\documentclass[letterpaper, 10 pt, conference]{ieeeconf}  
\usepackage{amsmath,amsfonts}
\usepackage{algorithmic}
\usepackage{algorithm}
\usepackage{array}
\usepackage[caption=false,font=normalsize,labelfont=sf,textfont=sf]{subfig}
\usepackage{textcomp}
\usepackage{stfloats}
\usepackage{url}
\usepackage{verbatim}
\usepackage{graphicx}
\usepackage{cite}
\usepackage{units}
\usepackage{dsfont}
\usepackage{comment}
\usepackage{booktabs}
\usepackage{marginnote}
\usepackage[dvipsnames]{xcolor}
\usepackage{xfrac}

\usepackage{amsthm}

\newtheorem{theorem}{Theorem}[section]

\newcounter{thm}
\newtheorem{prob}[thm]{Problem}

\IEEEoverridecommandlockouts                              

\newif\ifmargincomments 
\margincommentsfalse

\ifmargincomments
\newcommand{\obmargin}[2]{{\color{cyan}#1}\marginpar{\color{cyan}\raggedright\scriptsize [OB]:\\ #2}}

\else
\newcommand{\obmargin}[2]{#1}

\fi

\title{\LARGE \bf Effective Scaling of High-Fidelity Electric Motor Models for Electric Powertrain Design Optimization}

\author{Olaf Borsboom$^{1}$, Martijn Lokker$^{1}$, Mauro Salazar$^{1}$ and Theo Hofman$^{1}$
\thanks{This publication is part of the project NEON with project number 17628 of the research program Crossover, which is (partly) financed by the Dutch Research Council (NWO).}
\thanks{$^{1}$Department of Mechanical Engineering, Control System Technology,
        Eindhoven University of Technology, 5600 MB  Eindhoven, The Netherlands
        {\tt\small o.j.t.borsboom@tue.nl}}%
}

\begin{document}

\maketitle
\thispagestyle{empty}
\pagestyle{empty}

\begin{abstract}
In general, electric motor design procedures for automotive applications go through expensive trial-and-error processes or use simplified models that linearly stretch the efficiency map.
In this paper, we explore the possibility of efficiently optimizing the motor design directly, using high-fidelity simulation software and derivative-free optimization solvers.
In particular, we proportionally scale an already existing electric motor design in axial and radial direction, as well as the sizes of the magnets and slots separately, in commercial motor design software.
We encapsulate this motor model in a vehicle model together with the transmission, simulate a candidate design on a drive cycle, and find an optimum through a Bayesian optimization solver.
We showcase our framework on a small city car, and observe an energy consumption reduction of \unit[0.13]{\%} with respect to a completely proportional scaling method, with a motor that is equipped with relatively shorter but wider magnets and slots.
In the extended version of this paper, we include a comparison with the linear models, and add experiments on different drive cycles and vehicle types.
\end{abstract}

\section{Introduction}\label{sec:introduction}
In order to accelerate the adoption of electric vehicles (EVs), their purchasing price and driving range should be improved.
One strategy to achieve this, is to advance the design of the powertrain holistically.
However, the powertrain system is highly multidisciplinary and interconnected, and contains many design variables.
Focusing specifically on electric motors (EMs), which are complex machines, renders it difficult to optimize the component-level design with a system-level perspective.

One widely used method to optimize the size of EMs in powertrain design, is by scaling the EM, which can be achieved in multiple ways.
Two options are named here: The measurement data can be scaled along the torque axis, or the geometrical dimensions of an already existing design are scaled.
Both of these methods are built on very strong assumptions.
This calls for methods that EMs more accurately, but retain simplicity and tractability on the system level.

\subsubsection*{Related literature}
This paper addresses three closely connected research streams.
The first stream treats the optimization of (hybrid) EV powertrain sizing, such as in~\cite{BorsboomFahdzyanaEtAl2021,MurgovskiJohannessonEtAl2012,SilvasHofmanEtAl2016}.
In these cases, the sizing of the EM is carried out by linearly stretching the efficiency map along the torque axis.
However, this linear stretching method is considered to be valid only in small scalar ranges.

The second stream relates to the component-level design optimization of EMs, in order to, for instance, optimize total harmonic distortion or torque ripple \cite{LeiZhuEtAl2017,BramerdorferTapiaEtAl2018}.
The models that are used for this process are numerous in their design variables and finite-element methods, and thus viewed as fairly accurate, but this would be intractable for powertrain design purposes.

The final stream attempts to connect the first two streams: devising accurate EM models for powertrain design~\cite{RamakrishnanStipeticEtAl2018,StipeticGossEtAl2018,ClementeBorsboomEtAl2023}.
This is achieved by (re)scaling an already existing EM design in its geometrical dimensions, and then obtaining its performance by either simulating it with a higher fidelity tool, devising analytical scaling laws, or constructing surrogate models.
While all of these scaling methods are useful, they all consider proportional scaling of all design variables in the same direction with \textit{one} scaling factor.
This includes the internal design of the EM, which can be of great influence on the performance if considered separately.

Summarizing, to the best of the authors' knowledge, there are no methods for EM design optimization that consider internal design variables, while still retaining computational tractability on the powertrain system level.

\begin{figure}
	\centering
	\includegraphics[width=\columnwidth]{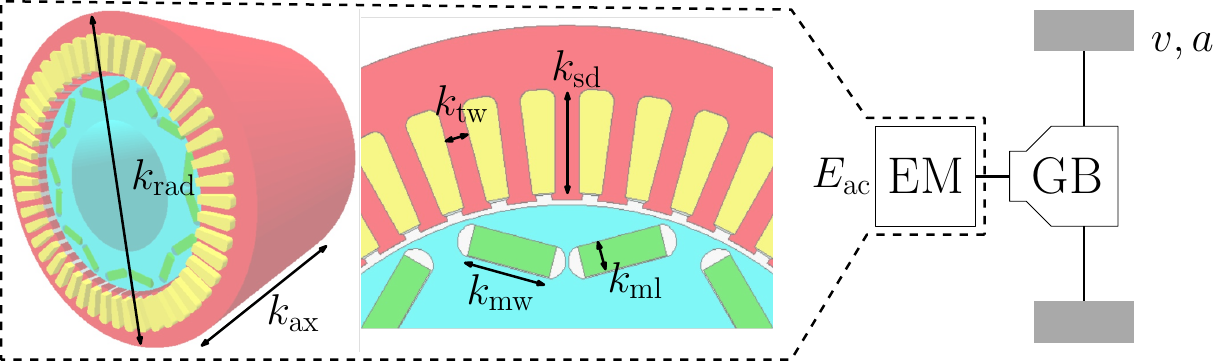}
	\caption{The simple powertrain, equipped with an electric motor (EM) and a transmission-final drive unit (GB) connected to the wheels. The EM is scaled proportionally in axial and radial direction, and internally in the magnets and slot sizes, denoted with scaling factors $k_{\{\cdot\}}$.}
	\label{fig:scaling}
\end{figure}%

\subsubsection*{Statement of Contributions}
In order to address this challenge, this paper contributes to the field by presenting a framework that optimizes the gear ratio value and the design of the EM in an electric powertrain directly, using high-fidelity simulations and derivative-free solvers. 
To this end, we scale the EM in its main geometric dimensions---axially and radially---and include the rescaling of the internal design variables related to the magnets and slots.
The scaling procedure is encapsulated in a powertrain and vehicle model, and showcased by optimizing the EM design for a small city car on a drive cycle.

\subsubsection*{Organization}
This paper is organized as follows:
Section~\ref{sec:methodology} presents the optimization framework, with a strong focus on the EM scaling models, and poses the optimization problem.
Section~\ref{sec:results} displays the results of showcasing the presented framework on a drive cycle, and the conclusions are drawn in Section~\ref{sec:conclusions}.

\section{Methodology}\label{sec:methodology}
We explain the optimization framework by introducing the vehicle and powertrain models, with emphasis on the scaling procedure of the EM (Section~\ref{subsec:longvehdyn} to \ref{subsec:em}).
Afterwards, we pose the optimization problem and solution approach in Section~\ref{subsec:optprob} and \ref{subsec:sol}, followed by a discussion in Section~\ref{subsec:discussion}.
For reasons of brevity, we will drop the notation of time dependency whenever it is clear from the context.

\subsection{Longitudinal Vehicle Dynamics}\label{subsec:longvehdyn}
We adopt a quasi-static modeling approach in this paper, as is common in powertrain design studies~\cite{GuzzellaSciarretta2007}.
Considering a flat road, we calculate the torque required at the wheels to complete the drive cycle $T_\mathrm{req}$ by
\begin{equation}
	T_\mathrm{req} = r_\mathrm{w} \cdot \left( m_\mathrm{v} \cdot a + \frac{1}{2}\cdot \rho_\mathrm{a} \cdot c_\mathrm{d} \cdot A_\mathrm{f} \cdot v^2 + c_\mathrm{r} \cdot m_\mathrm{v} \cdot g \right),
\end{equation}
where $v$ and $a$ are the drive cycle's velocity and acceleration, respectively, $r_\mathrm{w}$ is the radius of the wheels, $m_\mathrm{v}$ is the mass of the vehicle, $\rho_\mathrm{a}$ is the density of air, $c_\mathrm{d}$ is the aerodynamic drag coefficient, $A_\mathrm{f}$ is the frontal area of the vehicle, $c_\mathrm{r}$ is the rolling resistance coefficient, and $g$ is the gravitational constant.

\subsection{Transmission}\label{subsec:gb}
The one-speed transmission and final drive are modeled with a fixed efficiency $\eta_\mathrm{g}$ and gear ratio $\gamma$, which is subject to optimization within bounds $\underline{\gamma}$ and $\overline{\gamma}$.
Then, the torque at the EM shaft $T_\mathrm{m}$ can be calculated by
\begin{align}
	T_\mathrm{m} =
	\begin{cases}
		\frac{T_\mathrm{req}}{\gamma \cdot \eta_\mathrm{g}} \quad &\text{ if } T_\mathrm{req} \geq 0 \\
		\frac{T_\mathrm{req}\cdot \eta_\mathrm{g}}{\gamma} \quad &\text{ if } T_\mathrm{req} < 0,
	\end{cases} 
\end{align}
\obmargin{whereas}{regen braking fraction?} the EM speed $\omega_\mathrm{m}$ is given by
\begin{equation}
	\omega_\mathrm{m} = \gamma \cdot \frac{v}{r_\mathrm{w}}. 
\end{equation}

\subsection{Electric Motor}\label{subsec:em}
In this paper, we take inspiration from the \obmargin{proportional}{uniform? external?} scaling procedure for recalculation in~\cite{StipeticGossEtAl2018} of brushless permanent-magnet machines, without taking rewinding into consideration.
The proportional scaling procedure is shown in Fig.~\ref{fig:scaling} and described by two scaling factors: $k_\mathrm{ax} \in [\underline{k}_\mathrm{ax}, \overline{k}_\mathrm{ax}]$ and $k_\mathrm{rad}\in [\underline{k}_\mathrm{rad}, \overline{k}_\mathrm{rad}]$, which scale all design variables uniformly in the axial and radial direction, respectively, within the boundaries $\underline{k}_{\{\cdot\}}$ and $\overline{k}_{\{\cdot\}}$.
We adopt the same assumptions from~\cite{StipeticGossEtAl2018}, meaning that the cross-section of a \obmargin{proportionally}{} scaled EM is identical with the unscaled version, and the magnetic and electric loading remain unchanged~\cite{HendershotMiller2010}.

While all internal design parameters are altered with the proportional scaling factors, we add another layer of design freedom in our framework, where we allow the magnets and the slots to be rescaled separately in two dimensions each, on top of the proportional scaling factor.
In this way, we can change the magnetic and electric loading of the motor, which are the two primary elements that contribute to the torque per unit volume.
This can be viewed in Fig.~\ref{fig:scaling}, and the main equations are 
\begin{align}
	d_\mathrm{mw} &= d_\mathrm{mw,0} \cdot k_\mathrm{rad}\cdot k_\mathrm{mw}, \\
	d_\mathrm{ml} &= d_\mathrm{ml,0} \cdot k_\mathrm{rad}\cdot k_\mathrm{ml}, \\
	d_\mathrm{sd} &= d_\mathrm{sd,0} \cdot k_\mathrm{rad}\cdot k_\mathrm{sd}, \\
	d_\mathrm{tw} &= d_\mathrm{tw,0} \cdot k_\mathrm{rad}\cdot k_\mathrm{tw},
\end{align}
where $d_\mathrm{mw}$ and $d_\mathrm{ml}$ are the width and length of the magnets, respectively, $d_\mathrm{sd}$ is the depth of the slots, $d_\mathrm{tw}$ are the width of the teeth, $d_{\{\cdot\},0}$ are the dimensions of the unscaled motor, and all $k_{\{\cdot\}} \neq k_\mathrm{rad}$ are the individual rescaling factors of the internal design parameters with boundaries $\underline{k}_{\{\cdot\}}$ and $\overline{k}_{\{\cdot\}}$.

Each EM that is scaled proportionally and rescaled internally, a procedure that we will call \obmargin{\textit{combined scaling}}{??}, is simulated in Motor-CAD~\cite{MotorCAD}.
After the data on the operational limits and the losses $P_\mathrm{loss}$ are obtained, we calculate the power supplied to the motor $P_\mathrm{ac}$ as
\begin{equation}
	P_\mathrm{ac} =	T_\mathrm{m}\cdot \omega_\mathrm{m} + P_\mathrm{loss}\left(T_\mathrm{m}, \omega_\mathrm{m}\right).
\end{equation}
The energy used at the input of the EM $E_\mathrm{ac}$ is then computed by
\begin{equation}
	\frac{\mathrm{d}}{\mathrm{d}t} E_\mathrm{ac} = P_\mathrm{ac}.
\end{equation}

\subsection{Optimization Problem}\label{subsec:optprob}
The objective of the optimization problem in our framework is to minimize the used energy over the drive cycle $E_\mathrm{ac}(T)$, where $T$ is the duration of the cycle.
This is subject to the constraints on the design and input variables, and performance constraints.
The design variables $p$ are the scaling factors and the transmission ratio. 
The possible input variable $u$ is the selected gear.
The constraints on the performance comprise a desired maximum speed $v_\mathrm{max}$, acceleration time $t_\mathrm{acc}$ to reach velocity $v_\mathrm{acc}$, and gradeability on a slope $\alpha_{\mathrm{max}}$.

\begin{prob}[EM Design Problem]\label{prob:main}
	The optimal EM design is the solution of
	\begin{equation*}
		\begin{aligned}
			&\!\min & &E_\mathrm{ac}(T) \\
			& \textnormal{s.t. } & &\textnormal{Design Variable Constraints}\\
			& & &\textnormal{Input Variable Constraints}\\
			& & &\textnormal{Performance Constraints}.
		\end{aligned}
	\end{equation*}
\end{prob}

\subsection{Solution Approach}\label{subsec:sol}
\begin{figure}
	\centering
	\includegraphics[width=\columnwidth]{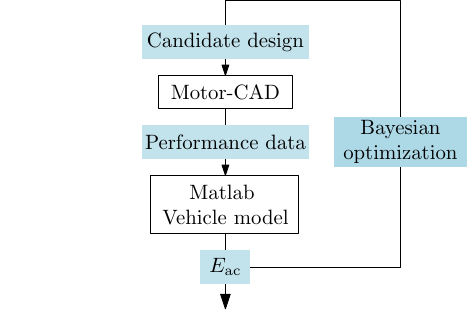}
	\caption{The solution approach of the design problem.}
	\label{fig:solstrategy}
\end{figure}

The approach to obtain a design solution is displayed in Fig.~\ref{fig:solstrategy}.
Because each candidate EM design is simulated in Motor-CAD, which is black-box and time-consuming, Bayesian optimization is adopted as the optimizer in Matlab.
The interface between Matlab and Motor-CAD is facilitated by ActiveX scripting.
After the data is obtained of each candidate design, the vehicle and its powertrain are simulated over the drive cycle.
When the optimizer has found a solution, the procedure is terminated.

\subsection{Discussion}\label{subsec:discussion}
A few comments are in order.
First, we choose to only consider continuous optimization variables in this paper.
Our framework could be extended by including integer design variables, such as the rewinding factor~\cite{StipeticGossEtAl2018}, and the number of poles pairs and slots, at the price of a longer computational time.
Second, since we select Bayesian optimization as the solver in our framework, we do not have any global optimality guarantees and our framework is not deterministic.
However, if we select sufficient optimization iterations, we are confident that the solver will have converged and we are provided with a promising design, from which EM experts can move forward to tune the design further, to yield higher efficiency, torque density, etc.
\obmargin{}{Willans line}


\section{Results}\label{sec:results}
In this section, we present \obmargin{the preliminary results}{bit insecure} of applying our framework to find an optimal EM design.
We use a template interior permanent magnet motor design in Motor-CAD as our reference.
This template is based on the motor designed for the 2012 Nissan Leaf, which is considered a compact hatchback type car, and is rated at \unit[120]{kW} of maximum power.
We employ our framework to design a new EM for a small city car, which is lighter than the Nissan Leaf reference vehicle, with both the proportional scaling and the combined scaling methods, on the Worldwide harmonized Light-duty vehicle Test Cycle (WLTC).
The vehicle parameters and constraint values are given in Table~\ref{tab:params}.
We allow the Bayesian optimization solver to run 50 iterations, where a Motor-CAD calculation of a scaled design and a drive cycle simulation take about \unit[220]{s} to evaluate on an ASUS VivoBook 15 notebook with an Intel i7-10750H 2.6 GHz CPU and 16 GB of RAM.
The total computation time amounts to an average of \unit[2]{h} and \unit[45]{min} for the proportional case (3 optimization variables) and \unit[3]{h} and \unit[45]{min} for the combined case (7 optimization variables).

\begin{small}
	\begin{table}
		\centering 
		\caption{Parameters} 
		\label{tab:params} 
		\begin{tabular}{c c c | c c c } 
			\toprule
			\bfseries Parameter & \bfseries Value & \bfseries Unit & \bfseries Parameter & \bfseries Value & \bfseries Unit\\ 
			\midrule 
			$m_\mathrm{v}$ & 1085 & \unit{kg}&					$r_\mathrm{w}$ & 0.295 & \unit{m} \\
			$A_\mathrm{f}$ & 0.72 & \unit{m$^2$} &				$c_\mathrm{d}$ & 0.35 & \unit{1} \\
			$g$ & 9.81 & \unit{m/s$^2$} &						$\rho$ & 1.2 & \unit{kg/m$^3$} \\		
			$\underline{k}_{\{\mathrm{ax,rad}\}}$ & 0.8 & \unit{-} & 		$\overline{k}_{\{\mathrm{ax,rad}\}}$ & 1.2 & \unit{-} \\
			$\underline{k}_{\{\mathrm{mw,ml,sd,tw}\}}$ & 0.9 & \unit{-} & 	$\overline{k}_{\{\mathrm{mw,ml,sd,tw}\}}$ & 1.1 & \unit{-} \\
			$\underline{\gamma}$ & 1 & \unit{-}& 				$\overline{\gamma}$ & 10 & \unit{-} \\
			$\eta_\mathrm{g}$ & 95 & \unit{\%}&					$r_\mathrm{w}$ & 0.295 & \unit{m} \\
			$v_\mathrm{max}$ & 180 & \unit{km/h} &				$v_{\mathrm{acc}}$ & 100 & \unit{km/h} \\
			$t_\mathrm{acc}$ & 9.6 & \unit{s} &					$\alpha_{\mathrm{max}}$ & 20 & \unit{\%} \\
			\bottomrule 
		\end{tabular} 
	\end{table} 
\end{small}

\subsection{Numerical Results}
The optimal designs that are obtained by our framework are summarized in Table~\ref{tab:solutions}.
For clarification, in the proportional case, the scaling factors related to the magnets and slots are not free variables in the optimization and are all fixed to have a value of 1.
Therefore they are excluded from this column in the table. 
As can be observed, the energy consumption reduction is fairly small: a difference of \unit[-0.13]{\%}.
However, the design solutions are significantly different, especially in the radial direction.
Moreover, in the combined scaling case, the slot and magnet sizes are wider but shorter than in the proportional one (note that a smaller $k_\mathrm{tw}$ refers to narrower teeth, and thus a wider slot).

\begin{small}
	\begin{table}
		\centering 
		\caption{Design Solutions} 
		\label{tab:solutions} 
		\begin{tabular}{c | c c } 
			\toprule
			\bfseries Solution & \bfseries Combined & \bfseries Proportional \\
			\midrule
			\bfseries $k_\mathrm{ax}$ & 0.81 & 0.80 \\
			\bfseries $k_\mathrm{rad}$ & 1.10 & 1.03 \\
			\bfseries $\gamma$ & 5.44 & 5.55 \\
			\bfseries $k_\mathrm{mw}$ & 1.05 & - \\
			\bfseries $k_\mathrm{ml}$ & 0.92 & - \\
			\bfseries $k_\mathrm{sd}$ & 0.95 & - \\
			\bfseries $k_\mathrm{tw}$ & 0.90 & - \\
			\bfseries $E_\mathrm{ac}(T)$ & \unit[7.81]{MJ} & \unit[7.82]{MJ} \\ 
			\bottomrule 
		\end{tabular} 
	\end{table} 
\end{small}

\section{Conclusions}\label{sec:conclusions}
In this paper, we instantiated a framework to optimize the size of electric motors in powertrains, by scaling already existing electric motor designs using high-fidelity tools.
We scale the motors in axial and radial direction, and add complexity by rescaling the sizes of the magnets and slots independently, giving the user more design freedom while maintaining a manageable number of design variables and tractability on system-level powertrain design optimization.
Our results show that, while the energy consumption reduction is fairly small, the framework does converge to a different optimum design w.r.t. the proportional scaling method.
This provides electric motor design engineers with multiple promising starting points for further design optimization.

In the full version of this paper, we aim to extend our experiments to optimize the motor design of a heavier cross-over type car, and to run on different drive cycles.
We also plan to focus on using our framework to validate the linearly stretching method described in Section~\ref{sec:introduction}, and the Willans line models~\cite{GuzzellaSciarretta2007}.
For future work, we would be interested in including integer design variables, such as the rewinding factors and the number of poles and slots.

\addtolength{\textheight}{0cm}   


\section*{Acknowledgment}

We thank Dr.~I.~New for proofreading this paper. We also thank Ir.~J.~Goudswaard for his input and expertise on electric motor design.



\bibliographystyle{IEEEtran}
\bibliography{../../../Bibliography/main,../../../Bibliography/SML_papers} 

\newcommand{\noopsort}[1]{} \newcommand{\printfirst}[2]{#1}
  \newcommand{\singleletter}[1]{#1} \newcommand{\switchargs}[2]{#2#1}
\begin{thebibliography}{10}
\providecommand{\url}[1]{#1}
\csname url@samestyle\endcsname
\providecommand{\newblock}{\relax}
\providecommand{\bibinfo}[2]{#2}
\providecommand{\BIBentrySTDinterwordspacing}{\spaceskip=0pt\relax}
\providecommand{\BIBentryALTinterwordstretchfactor}{4}
\providecommand{\BIBentryALTinterwordspacing}{\spaceskip=\fontdimen2\font plus
\BIBentryALTinterwordstretchfactor\fontdimen3\font minus
  \fontdimen4\font\relax}
\providecommand{\BIBforeignlanguage}[2]{{%
\expandafter\ifx\csname l@#1\endcsname\relax
\typeout{** WARNING: IEEEtran.bst: No hyphenation pattern has been}%
\typeout{** loaded for the language `#1'. Using the pattern for}%
\typeout{** the default language instead.}%
\else
\language=\csname l@#1\endcsname
\fi
#2}}
\providecommand{\BIBdecl}{\relax}
\BIBdecl

\bibitem{BorsboomFahdzyanaEtAl2021}
O.~Borsboom, C.~A. Fahdzyana, T.~Hofman, and M.~Salazar, ``A convex
  optimization framework for minimum lap time design and control of electric
  race cars,'' \emph{{IEEE Transactions on Vehicular Technology}}, vol.~70,
  no.~9, pp. 8478--8489, 2021.

\bibitem{MurgovskiJohannessonEtAl2012}
N.~Murgovski, L.~Johannesson, J.~Sj\"oberg, and B.~Egardt, ``Component sizing
  of a plug-in hybrid electric powertrain via convex optimization,''
  \emph{Mechatronics}, vol.~22, no.~1, pp. 106--120, 2012.

\bibitem{SilvasHofmanEtAl2016}
E.~Silvas, T.~Hofman, N.~Murgovski, P.~Etman, and M.~Steinbuch, ``Review of
  optimization strategies for system-level design in hybrid electric
  vehicles,'' \emph{{IEEE Transactions on Vehicular Technology}}, vol.~66,
  no.~1, pp. 57--70, 2016.

\bibitem{LeiZhuEtAl2017}
G.~Lei, J.~Zhu, Y.~Gou, C.~Liu, and B.~Ma, ``A review of design optimization
  methods for electrical machines,'' \emph{{Energies}}, vol.~10, no.~12, 2017.

\bibitem{BramerdorferTapiaEtAl2018}
G.~Bramerdorfer, J.~Tapia, J.~J. Pyrhönen, and A.~Cavagnino, ``Modern
  electrical machine design optimization: Techniques, trends, and best
  practices,'' \emph{{IEEE Transactions on Industrial Electronics}}, vol.~65,
  no.~10, pp. 7672--7684, 2018.

\bibitem{RamakrishnanStipeticEtAl2018}
K.~Ramakrishnan, S.~Stipetic, M.~Gobbi, and G.~Mastinu, ``Optimal sizing of
  traction motors using scalable electric machine model,'' \emph{{IEEE
  Transactions on Transportation Electrification}}, vol.~4, no.~1, pp.
  314--321, 2018.

\bibitem{StipeticGossEtAl2018}
S.~Stipetic, J.~Goss, J.~Zarko, and M.~Popescu, ``Calculation of efficiency
  maps using a scalable saturated model of synchronous permanent magnet
  machines,'' \emph{{IEEE Transactions on Industry Applications}}, vol.~54,
  no.~5, pp. 4257--4267, 2018.

\bibitem{ClementeBorsboomEtAl2023}
M.~Clemente, O.~Borsboom, M.~Salazar, and T.~Hofman, ``A geometric {EM} model
  for optimal vehicle family design,'' in \emph{{IEEE Conf.\ on Control
  Technology and Applications}}, 2023, under review.

\bibitem{GuzzellaSciarretta2007}
L.~Guzzella and A.~Sciarretta, \emph{Vehicle propulsion systems: Introduction
  to Modeling and Optimization}, 2nd~ed.\hskip 1em plus 0.5em minus 0.4em\relax
  {Springer Berlin Heidelberg}, 2007.

\bibitem{HendershotMiller2010}
J.~R. Hendershot and T.~J.~E. Miller, \emph{Design of Brushless
  Permanent-Magnet Machines}, 1st~ed.\hskip 1em plus 0.5em minus 0.4em\relax
  {Motor Design Books}, 2010.

\bibitem{MotorCAD}
{Ansys Motor-CAD}. ANSYS, Inc. {Available at}
  \url{https://www.ansys.com/products/electronics/ansys-motor-cad}.

\end{thebibliography}

\end{document}